\documentclass[aps,prb,amsmath,amssymb,floatfix,twocolumn,amsmath,superscriptaddress,twocolumn,nofootinbib,tighten,letterpaper]{revtex4-2}

\usepackage[colorlinks,linkcolor=blue,citecolor=blue,urlcolor=blue]{hyperref}

\usepackage{multirow}
\usepackage{subfigure}
\usepackage{color}
\usepackage{mathrsfs}
\usepackage{hyperref}
\usepackage[normalem]{ulem}
\usepackage{bm}

\usepackage{amssymb}   
\usepackage{amsmath}
\renewcommand\vec[1]{\ensuremath\boldsymbol{#1}} 

\usepackage{amsfonts, relsize, color}
\usepackage{graphics}
\usepackage{graphicx}
\usepackage{subfigure}
\usepackage{hyperref}
\usepackage{color}
\usepackage{comment}

\begin{document}

\title{Dynamic mass generation on two-dimensional electronic hyperbolic lattices}

\author{Noble Gluscevich}
\altaffiliation{These authors contributed equally to this work.}
\affiliation{Department of Physics, Lehigh University, Bethlehem, Pennsylvania, 18015, USA}

\author{Abhisek Samanta}
\altaffiliation{These authors contributed equally to this work.}
\affiliation{Department of Physics, The Ohio State University, Columbus, Ohio 43210, USA}
\affiliation{Department of Physics, Indian Institute of Technology Gandhinagar, Gujarat 382355, India}

\author{Sourav Manna}
\affiliation{Department of Condensed Matter Physics, Weizmann Institute of Science, Rehovot 7610001, Israel}
\affiliation{Raymond and Beverly Sackler School of Physics and Astronomy, Tel-Aviv University, Tel Aviv 6997801, Israel}

\author{Bitan Roy}~\thanks{Contact author: bitan.roy@lehigh.edu}
\affiliation{Department of Physics, Lehigh University, Bethlehem, Pennsylvania, 18015, USA}

\begin{abstract}
Free electrons hopping on hyperbolic lattices embedded on a negatively curved space can foster (a) Dirac liquids, (b) Fermi liquids, and (c) flat bands, respectively characterized by a vanishing, constant, and divergent density of states near the half filling. From numerical self-consistent mean-field Hartree analyses, we show that nearest-neighbor Coulomb and on-site Hubbard repulsions respectively give rise to charge-density-wave and antiferromagnetic orders featuring staggered patterns of average electronic density and magnetization in all these systems, when the hyperbolic tessellation is accomplished by periodic arrangements of even $p$-gons. Both quantum orders dynamically open mass gaps near the charge neutrality point via spontaneous symmetry breaking. Only on hyperbolic Dirac materials these orderings take place via quantum phase transitions (QPTs) beyond critical interactions, which however decrease with increasing curvature, showcasing curvature-induced weak-coupling QPTs. We present scaling of these masses with the corresponding interaction strengths. 
\end{abstract}

\maketitle

\emph{Introduction}.~From table salts to quantum materials, crystals are ubiquitous in nature. They feature discrete translational, rotational and reflection symmetries, manifesting periodic arrangements of regular polygons with $p$ arms ($p$-gon). Yet another integer plays a pivotal role for the geometric classification of crystals, the number of nearest-neighbor (NN) sites $q$, connected to each vertex of a $p$-gon also known as the coordination number. On Euclidean plane, these symmetry constraints lead to $(p-2)(q-2)=4$, permitting triangular ($2p=q=6$), square ($p=q=4$), and honeycomb ($p=2q=6$) lattices. By contrast, hyperbolic tessellation on a negatively curved space, is accomplished when an inequality $(p-2)(q-2)>4$ is satisfied. Therefore, hyperbolic space accommodates infinitely many periodic lattices, characterized by the Schl\"afli symbol $(p,q)$~\cite{TarjusJPhA2007, KollarFitzpatrick2019, AsaduzzamanCatterallPhysRevD2020, BoettcherGorshkov2020PRA, BrowerCogburnPhysRevD2021, MaciejkoRayanSciAdv2021,  IkedaAokiJournalofPhysics2021, BoettcherGorshkovPhysRevB2022, ChengSerafinPhysRevLett2022, MaciejkoRayan2022, AttarBoettcherPhysRevE2022, MosseriVogelerPhysRevB2022, BzdusekMaciejkoPhysRevB2022, KienzleRayan2022, StegmaierUpretiPhysRevLett2022}.

\begin{figure*}[t!]
\includegraphics[width=1.00\linewidth]{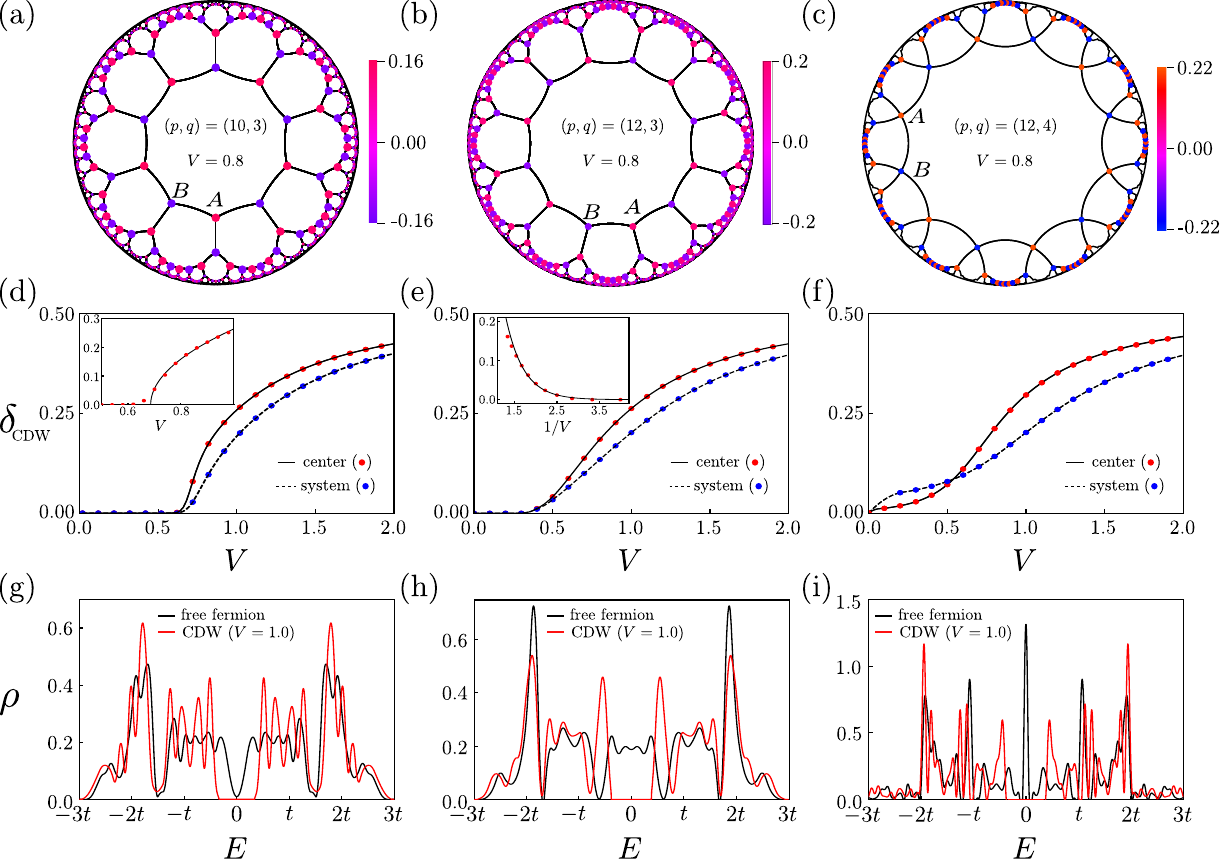}
\caption{CDW on HLs (shown on a Poincar\'e disk) resulting from the NN Coulomb repulsion ($V$). Spatial variation of the average electronic density measured from the half-filling on (a) $(10,3)$, (b) $(12,3)$, and (c) $(12,4)$ HLs for $V=0.8$. Scaling of the CDW order parameter at the center, constituted by the sites belonging to the zeroth generation or center $p$-gon of the system~\cite{comment1} (red) and averaged over the entire system (blue) with $V$ are shown in (d)-(f), respectively. DOS for free fermions (black) and with the CDW order for $V=1.0$ (red) are shown in (g)-(i), respectively, displaying the formation of a mass gap. Here, the results are presented for third generation $(10,3)$, $(12,3)$, and $(12,4)$ HLs, respectively containing $2880$, $7680$, and $13080$ sites~\cite{comment1}. Critical coupling for the CDW ordering in a $(10,3)$ Dirac system is estimated to be $V_c \approx 0.69$ [inset of (d)], while in $(12,3)$ Fermi liquids it follows the BCS scaling [inset of (e)]. 
}~\label{fig:FigCDW}
\end{figure*}

\begin{figure*}[t!]
\includegraphics[width=1.00\linewidth]{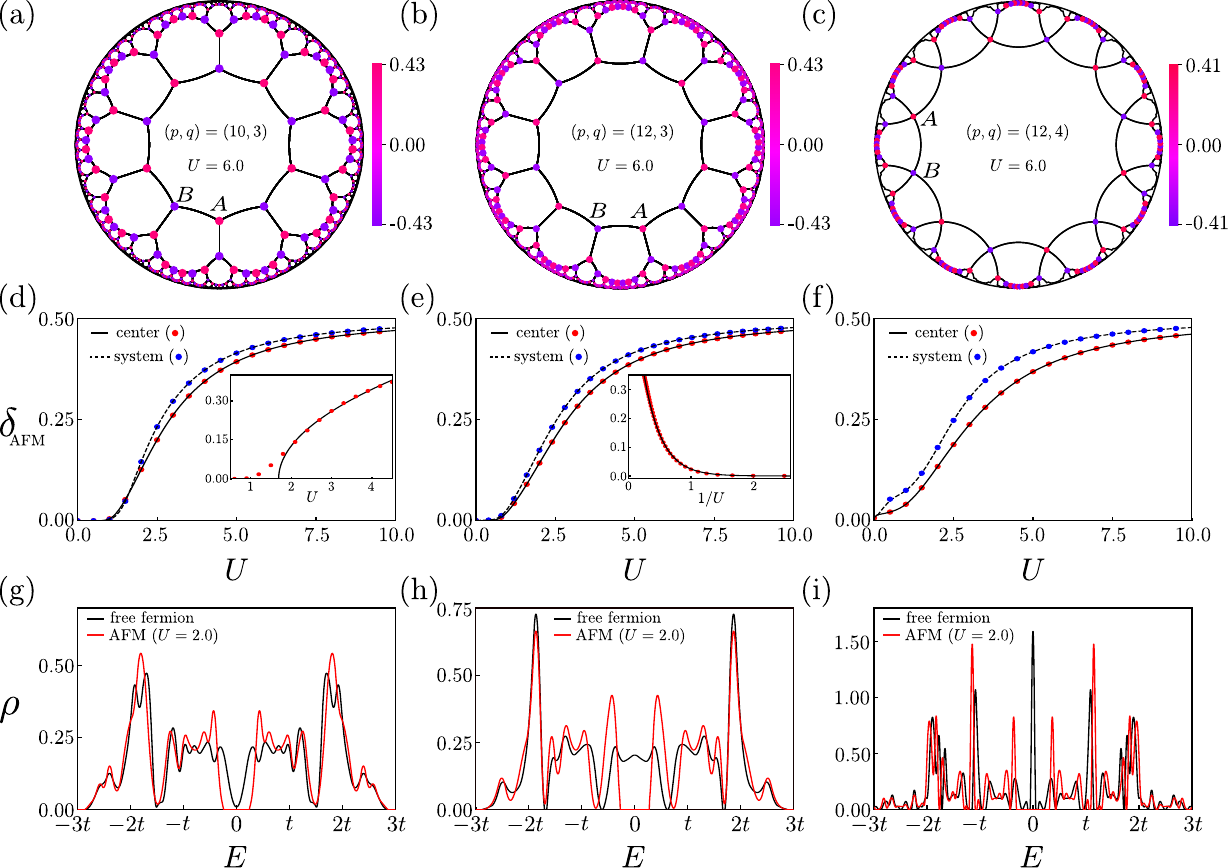}
\caption{On-site Hubbard repulsion ($U$) mediated AFM ordering [Eq.~\eqref{eq:AFMOP}] on HLs (shown on a Poincar\'e disk). Top row: Self-consistent solutions of magnetization at each site measured from its value at half filling (zero), showing AFM order for $U=6.0$. The results are presented for third generation $(10,3)$, and second generation $(12,3)$ and $(12,4)$ HLs, respectively containing $2880$, $972$, and $1320$ sites. All the other details are the same as in Fig.~\ref{fig:FigCDW}. The ``center" is constituted by the sites from the zeroth generation or center $p$-gon of the system~\cite{comment1}.        
}~\label{fig:FigAFM}
\end{figure*}

Naturally, hyperbolic quantum materials harbor a variety of electronic band structures~\cite{MaciejkoRayanSciAdv2021, IkedaAokiJournalofPhysics2021, BoettcherGorshkovPhysRevB2022, ChengSerafinPhysRevLett2022, MaciejkoRayan2022, AttarBoettcherPhysRevE2022, MosseriVogelerPhysRevB2022, BzdusekMaciejkoPhysRevB2022, KienzleRayan2022}. They can be classified in terms of the density of states (DOS) near half-filling or zero-energy. A family of hyperbolic lattices (HLs) with even integer $p$, constituting bipartite lattices with the NN sites belonging to $A$ and $B$ sublattices [Figs.~\ref{fig:FigCDW} and~\ref{fig:FigAFM}], fosters (a) Dirac liquids, (b) Fermi liquids, and (c) flat bands depending on $p$ and $q$, captured from a free-electron tight-binding model with only the NN hopping between sites from $A$ and $B$ sublattices. Respectively, they feature vanishing, constant, and divergent DOS at zero energy [$\rho(0)$]. See Table~\ref{tab:HLclassification}.   

\begin{table}[b!]
\begin{tabular}{|cc|cc|cc|cc|}
\hline
\multicolumn{2}{|c|}{$p$-gon} & \multicolumn{2}{c|}{Dirac liquid} & \multicolumn{2}{c|}{Fermi liquid} & \multicolumn{2}{c|}{Flat band} \\ \hline \hline
\multicolumn{2}{|c|}{$p/2=$ odd} & \multicolumn{2}{c|}{$q=3$} & \multicolumn{2}{c|}{$q=4,5$} & \multicolumn{2}{c|}{$\times$} \\ \hline
\multicolumn{2}{|c|}{$p/2=$ even} & \multicolumn{2}{c|}{$\times$} & \multicolumn{2}{c|}{$q=3$} & \multicolumn{2}{c|}{$q=4,5$} \\ \hline
\end{tabular}
\caption{Classification of electronic bipartite (even $p$) HLs in terms of the DOS at zero energy $\rho(0)$. The Dirac liquid, Fermi liquid, and flat band are identified from vanishing, constant, and divergent $\rho(0)$, respectively.  
}~\label{tab:HLclassification}
\end{table}

\begin{figure}[t!]
\includegraphics[width=1.00\linewidth]{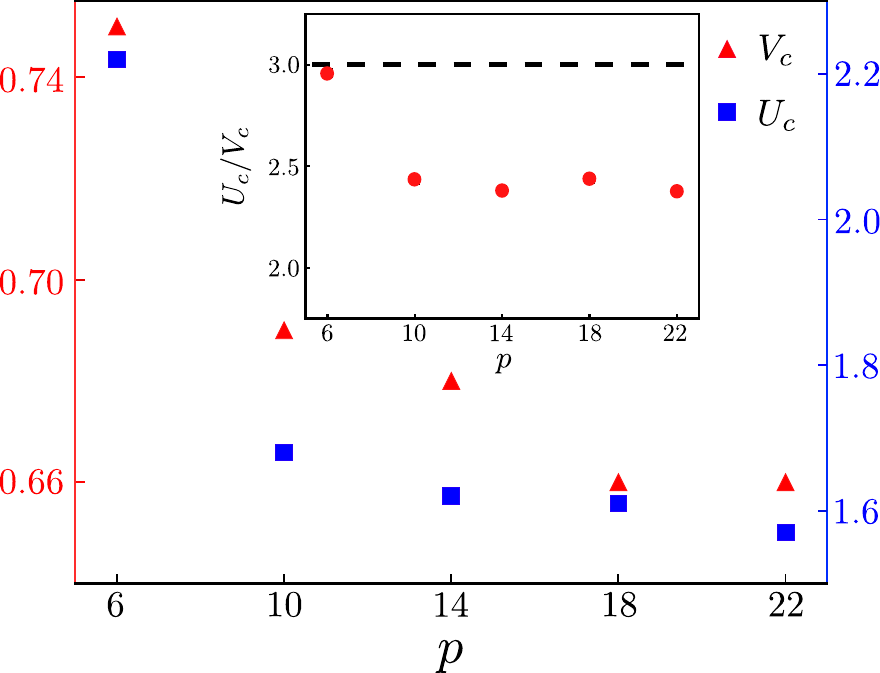}
\caption{Critical NN ($V_c$) and on-site ($U_c$) interactions for the CDW and AFM orderings, respectively, on graphene or Euclidean ($p=6$) and hyperbolic ($p=10, 14,18,22$) Dirac materials ($q=3$), displaying their reduction with increasing $p$ or curvature, indicating QPTs at weaker coupling on curved space. Inset: $U_c/V_c$ in Dirac materials, expected to be locked at $3$ (dashed line) in the mean-field limit. 
}~\label{fig:Figcritical}
\end{figure}

On such electronic HLs with an open boundary condition (OBC), we numerically investigate the role of NN Coulomb ($V$) and on-site Hubbard ($U$) repulsions at half filling within the mean-field or Hartree approximation. Respectively, they support staggered patterns of electronic density and spin, two quantum phases named the charge density wave (CDW) and antiferromagnet (AFM). While infinitesimal $V$ and $U$ are sufficient to nucleate them on HLs hosting Fermi liquids and flat bands, due to the linearly vanishing DOS such orderings take place at finite interactions via quantum phase transitions (QPTs) in Dirac liquids. See Figs.~\ref{fig:FigCDW} and~\ref{fig:FigAFM}. More intriguingly, as the curvature is strengthened in Dirac materials (realized for larger $p$ with $q=3$), the critical couplings for these two ordering decrease monotonically, indicating curvature-induced weak-coupling QPTs therein. See Fig.~\ref{fig:Figcritical}.

\emph{Free fermions}.~A tight-binding model of free fermions, hopping only between the NN sites of HLs, is given by 
\begin{equation}
H_0 =-\sum_{\langle i, j \rangle} t_{ij} c^\dagger_i c_j,  
\end{equation}    
where $c^\dagger_i$ ($c_i$) is the fermionic creation (annihilation) operator on the $i$th site and $\langle \cdots \rangle$ restricts the summation within the NN sites~\cite{comment1}. The spin independent NN hopping amplitude $t_{ij}$ is assumed to be constant $t$, which we set to be unity. For simplicity, we suppress the fermionic spin index. Inclusion of spin only causes a mere doubling of $H_0$. The Hamiltonian operator ($\hat{h}_0$) associated to $H_0$ preserves (a) time-reversal symmetry (${\mathcal T}$), ${\mathcal T} \hat{h}_0 {\mathcal T}^{-1}=+\hat{h}_0$ with ${\mathcal T}^2=+1$, (b) anti-unitary particle-hole symmetry (${\mathcal P}$), ${\mathcal P} \hat{h}_0 {\mathcal P}^{-1}=-\hat{h}_0$ with ${\mathcal P}^2=+1$, and (c) unitary chiral or sublattice symmetry (${\mathcal C}$), ${\mathcal C} \hat{h}_0 {\mathcal C}^{-1}=-\hat{h}_0$ with ${\mathcal C}^2=1$. Hence, the system belongs to the class BDI~\cite{SchnyderRyuFurusakiLudwig2008, SM}. A half-filled system keeps all the negative (positive) energy states occupied (empty), and the average fermionic density at each site is $1/2$, manifesting the sublattice exchange symmetry. Then $\rho(0)$ serves as a smoking gun probe to group quantum hyperbolic materials into three broad classes, characterized by vanishing (Dirac liquid), constant (Fermi liquid), and divergent (flat band) $\rho(0)$. See Table~\ref{tab:HLclassification}. A slight deviation from a perfect $|E|$-linear DOS in hyperbolic Dirac materials very close to $E=0$ results from finite-size effects. DOS computed from hyperbolic band theory~\cite{BoettcherGorshkovPhysRevB2022} shows a perfect $|E|$-linear behavior near $E=0$~\cite{SM}.

\emph{NN repulsion}.~We now investigate the role of electronic interactions in these three families of quantum fluids, first considering spinless fermions and NN Coulomb repulsion ($V$) among them, captured by the Hamiltonian 
\begin{equation}
H_V = H_0 + \frac{V}{2} \sum_{\langle i,j \rangle} n_i n_j - \mu N.
\end{equation}
Here, $n_i=c^\dagger_i c_i$ is the fermionic density on site $i$, $N$ is the total number of spinless fermions in the half-filled system, and $\mu$ is the chemical potential. Performing a Hartree decomposition of the quartic term, we arrive at the effective single-particle Hamiltonian~\cite{RoyHerbutPhysRevB2011, RoySau2014, RoyPhysRevB2017} 
\begin{equation}
H^{\rm Har}_V= H_0 + V \sum_{\langle i,j \rangle} \bigg( \langle n_{B,i} \rangle n_{A, j} + \langle n_{A,i} \rangle n_{B, j} \bigg) - \mu N.
\end{equation}
Here, $\langle n_{A} \rangle$ ($\langle n_{B} \rangle$) corresponds to the site-dependent self-consistent average fermionic density on the sublattice $A$ ($B$). We measure them relative to the uniform density at half filling: $\langle n_{A,i}\rangle = 1/2 + \delta_{A,i}$ and $\langle n_{B,i}\rangle = 1/2 - \delta_{B,i}$. Half filling is maintained by choosing $\mu=V/2$ and ensuring that $\sum_{i} \left( \delta_{A,i} -\delta_{B,i} \right) =0$. Positive definite quantities $\delta_A$ and $\delta_B$ yield the local CDW order parameter 
\begin{equation}~\label{eq:CDWOP}
\delta_{\rm CDW}=\frac{1}{2} \left( \delta_{A} + \delta_{B} \right), 
\end{equation}
where for a given $A$ site, the $B$ site is one of the two NN sites belonging to the same generation of the hyperbolic lattice~\cite{comment1}. We numerically compute $\delta_{A}$ and $\delta_{B}$, and subsequently $\delta_{\rm CDW}$ in the entire system with OBC for a wide range of $V$. See Fig.~\ref{fig:FigCDW}.

When $\delta_A$ and $\delta_B$, thus $\delta_{\rm CDW}$ are finite in the entire system (not necessarily uniform), it becomes an insulator at half filling by spontaneously breaking the sublattice symmetry. To appreciate this outcome, we define a two-component superspinor $\Psi^\top=(c_A, c_B)$, where $c_A$ ($c_B$) is an $N$-dimensional spinor constituted by the annihilation operator on the sites from $A$ ($B$) sublattice. In this basis, the tight-binding Hamiltonian and the CDW order-parameter are respectively 
\begin{equation}~\label{eq:effectivehamil}
\hat{h}_0= \left( \begin{array}{cc}
{\boldsymbol 0} & {\bf t} \\
{\bf t}^\top & {\bf 0}
\end{array}
\right)
\:\: \text{and} \:\:
\hat{h}_{\rm CDW} =\left( \begin{array}{cc}
{\boldsymbol \Delta} & {\boldsymbol 0} \\
{\boldsymbol 0} & -{\boldsymbol \Delta}
\end{array} 
\right) \equiv \hat{h}_{\Delta},
\end{equation}    
where ${\boldsymbol 0}$ is an $N$-dimensional null matrix, ${\bf t}$ is the intersublattice hopping matrix, and $\pm {\boldsymbol \Delta}$ are $N$-dimensional diagonal matrices, whose entries are the self-consistent solutions of $\delta_A$ and $\delta_B$ at various sites of the system, respectively, and $\top$ denotes transposition. As $\hat{h}_0$ and $\hat{h}_{\rm CDW}$ \emph{anticommute} $\{ \hat{h}_0, \hat{h}_{\rm CDW} \}=0$, the CDW order acts as a mass for gapless fermions. Its spontaneous nucleation causes insulation near the charge neutrality point by opening a mass gap in the electronic spectra with vanishing [Fig.~\ref{fig:FigCDW}(g)], constant [Fig.~\ref{fig:FigCDW}(h)], and divergent [Fig.~\ref{fig:FigCDW}(i)] DOS at zero energy.

As $\rho (E) \sim |E|$ near the zero energy, condensation of the CDW on hyperbolic Dirac systems occurs beyond a critical  NN repulsion ($V_c$), slightly away from which $\delta_{\rm CDW} \sim \sqrt{V-V_c}$ for $V>V_c$, allowing us to estimate $V_c$ by minimizing the finite-size effects [inset of Fig.~\ref{fig:FigCDW}(d)]. By contrast, when $\rho(0)$ is finite or divergent, the CDW order sets in for infinitesimal $V$. In the former system, CDW follows the BCS scaling $\delta_{\rm CDW} \sim \exp(-\kappa/V)$ due to a constant $\rho (0)$, further substantiated from the exponential fit between $\delta_{\rm CDW}$ and $1/V$ [inset of Fig.~\ref{fig:FigCDW}(e)]. On a $(12,3)$ HL, the fitting parameter $\kappa=2.36$. In the Supplemental Material (SM) we justify these scaling behaviors from the solutions of a gap equation~\cite{RoyPhysRevB2017, Tinkham2004, SM}. When $\rho(0)$ is divergent, $\delta_{\rm CDW}$ grows considerably faster for weak $V$. Similar scaling holds on HLs with different $(p \;({\rm even}),q)$~\cite{SM}. With OBCs, $\delta_{\rm CDW}$ is not uniform, rather displaying spatial modulations in the system [Fig.~\ref{fig:FigCDW}(a)-(c)]. However, a uniform CDW order with $\delta_A=\delta_B=0.5$ develops in the entire system when $V \gg 1$. In this limit, spinless fermions occupy one sublattice of bipartite HLs, while the other one is empty.

\emph{Hubbard model}.~The Hubbard Hamiltonian with on-site repulsion ($U$) among spinful fermions (electrons) reads
\begin{equation}
H_U= U \sum_{i} \left( n_{i, \uparrow} -\frac{1}{2} \right) \; \left( n_{i, \downarrow} -\frac{1}{2} \right) -\mu N,
\end{equation} 
where $n_{i,\uparrow/\downarrow}$ is the electronic density at site $i$ with spin projection $\uparrow/\downarrow$ in the $z$ direction. The Hartree decomposition of $H_U$ leads to 
\begin{eqnarray}
H^{\rm Har}_{U} &=& \sum_{x=A,B} \bigg\{ \left( \langle n_{x,\uparrow} \rangle -\frac{1}{2}\right) \left( n_{x,\downarrow} -\frac{1}{2} \right) \nonumber \\
&+& \left( \langle n_{x,\downarrow} \rangle -\frac{1}{2}\right) \left( n_{x,\uparrow} -\frac{1}{2} \right) \bigg\} -\mu N,
\end{eqnarray}
where $\langle n_{A,\sigma} \rangle= 1/2 +\sigma \delta_{A,\sigma}(\vec{r})$, $\langle n_{B,\sigma} \rangle= 1/2 -\sigma \delta_{B,\sigma}(\vec{r})$, $\sigma=+/- \equiv \uparrow/\downarrow$, and $\vec{r}$ measures the position of a site~\cite{RoyAssaadPhysRevX2014, RoyPhysRevB2017}. Charge neutrality demands $\mu=0$ and
\begin{equation}
\sum_{\sigma=\pm} \; \sum_{\vec{r}} \; \sigma \; \left[ \delta_{A,\sigma} (\vec{r}) - \delta_{B, \sigma} (\vec{r}) \right]=0,
\end{equation}
which also keeps the average electronic density at each site equal to one. Then, an AFM ground state with $\delta_{A/B, \uparrow/\downarrow}>0$ is characterized by the order parameter
\begin{equation}~\label{eq:AFMOP}
\delta_{\rm AFM}=\frac{1}{2} \left( \delta_{A,\uparrow} + \delta_{A,\downarrow} + \delta_{B,\uparrow} + \delta_{B,\downarrow} \right),
\end{equation} 
where the $B$ site is one of the two NN sites in the same generation of the hyperbolic lattice for a given site belonging to the $A$ sublattice~\cite{comment1}. In noninteracting half-filled systems $\delta_{\rm AFM}=0$. We follow the same numerical procedure to self-consistently compute $\delta_{\rm AFM}$ everywhere in the system, outlined in detail for the CDW order. The results are summarized in Fig.~\ref{fig:FigAFM}.

Inside the AFM phase, each site develops a finite magnetization, pointing in the opposite directions on two complementary sublattices. Consequently, the net magnetization in the system is zero. The AFM order also develops a mass gap near the half-filling, which can be appreciated in the following way. For spinful fermions, the tight-binding Hamiltonian $\hat{h}^{\rm spin}_0=\sigma_0 \otimes \hat{h}_0$ and AFM order $\hat{h}_{\rm AFM}=\sigma_3 \otimes \hat{h}_\Delta$ anticommute, $\{ \hat{h}^{\rm spin}_0, \hat{h}_{\rm AFM} \}=0$ [Eq.~\eqref{eq:effectivehamil}]. Two-component Pauli matrices $\{ \sigma_\mu \}$ operate on the spin indices. Hence, the AFM order also acts as a mass for gapless fermions and its nucleation causes insulation at half-filling. Scaling of the AFM order and its spatial variations are qualitatively similar to the ones for CDW. For instance, critical on-site repulsion for AFM order is $U_c = 1.68$ on $(10,3)$ and the BCS fitting parameter $\kappa = 3.54$ on $(12,3)$ HLs, respectively. And on $(12,4)$ HL the AFM develops rapidly for weak $U$.

\emph{Weak-coupling QPT}.~The critical couplings for CDW and AFM orders on hyperbolic Dirac materials ($q=3$) decrease monotonically with increasing $p$ or the curvature, compared to the honeycomb lattice, harboring Dirac fermions on Euclidean flat land (Fig.~\ref{fig:Figcritical}), where $V_c \approx 0.75$~\cite{WeeksFranzPhysRevB2010, RoyHerbutPhysRevB2011} and $U_c \approx 2.22$~\cite{Sorella1992}, which we also reproduce. It strongly indicates curvature-induced QPTs at weaker coupling for Dirac fermions on curved space. Within the mean-field approximation $U_c/V_c$ is expected to be $3$, as the NN repulsion is operative between the fermionic densities on three pairs of NN sites, while the Hubbard repulsion is on-site. On hyperbolic lattices with OBC although $U_c/V_c \approx 2.4$ remains approximately constant, it is smaller than $3$ [Fig.~\ref{fig:Figcritical} (inset)], possibly due to a rapidly increasing large number of sites at the boundary of the system with only two NN sites, increasing $V_c$.

\emph{Summary and discussion}.~From numerical Hartree self-consistent analyses, we show that NN Coulomb and on-site Hubbard repulsions respectively support CDW and AFM orders on half-filled bipartite HLs. While in Dirac systems such orderings take place through a QPT at finite interactions, they nucleate even for infinitesimal $V$ and $U$ when the DOS at half-filling is either finite or divergent. The Fock term for the NN interaction $\langle c^\dagger_i c_j \rangle$ renormalizes the NN hopping amplitude, leaving the system gapless. As such any other decomposition of these interactions yields only gapless states, which are energetically inferior to the fully and isotropic gapped CDW and AFM states at zero temperature (no competition with entropy). From (a) the generation dependence of the local CDW and AFM orders on HLs with different system sizes, and (b) local DOS at its various generations, we show that the dynamic symmetry breaking takes place everywhere in the system and the entire system then becomes a correlated insulator~\cite{SM}. From numerical solutions of the Dyson equation~\cite{FetterWalecka}, we show that mean-field solutions for the CDW are stable against Gaussian fluctuations around it~\cite{SM}, which can be generalized for the amplitude mode of the AFM order, leaving aside its two gapless Goldstone modes. Due to a genuine two-dimensional (2D) nature of the HLs, in which the ratio of edge to bulk sites reaches a finite number~\cite{SunkyuPRL2020}, the Goldstone modes in the AFM order do not destroy any long-range order at zero temperature. Due to the two-dimensional nature of the hyperbolic lattices, the gravitational anomaly is absent therein and it does not impede our outcomes~\cite{FujikawaAnomalybook}. Also, on hyperbolic Dirac materials the critical interactions for these two orders decrease with increasing curvature or $p$. These predictions can be further tested from quantum Monte Carlo simulations, accounting for quantum fluctuations, as shown for Euclidean Dirac systems (graphene) with on-site~\cite{HerbutAssaadPhysRevX2013, SorellaPhysRevX2016} and NN~\cite{Troyer2014NJPhNNRepulsion, Yao2015NJPhNNRepulsion} repulsions. The universality class of the QPTs to CDW and AFM phases cab be captured by Gross-Neveu~\cite{GrossNeveu} or Nambu-Jona-Lasinio~\cite{NJL1} models for dynamic mass generation on curved space~\cite{curvedspaceRG1}, endowing a unique opportunity to test the predictions of curved space quantum field theory directly from HL-based numerical investigations. We note that the application of strong magnetic fields on hyperbolic Dirac materials by virtue of generating a finite $\rho(0)$ causes insulation in the half-filled system through CDW and AFM orderings even for infinitesimal $V$ and $U$, respectively~\cite{roycatalysishyperbolic}. In the future, we will search for other exotic correlated quantum phases on HLs, among which topological Mott insulators and superconductors are the fascinating ones.

Designer electronic materials~\cite{ManoharanNat2012, ManoharanReview2013, GomesNatCom2017, MoraisSmithNatPh2019, MoraisSmithNatMat2019} and cold atomic setups are promising platforms where our predicted quantum phases can be observed experimentally. A curved designer material can be engineered by growing its substrate (typically Cu) on another material with a different thermal expansion coefficient. When such a heterostructure is cooled down, the Cu substrate gets curved, which then can be decorated by the sites of desired HLs. With tunable hopping amplitudes $t$, the ratios $V/t$ and $U/t$ can be adjusted on hyperbolic designer quantum materials to trigger various quantum orders. In cold atomic setups, desired hyperbolic tessellations can be achieved by suitable arrangements of laser traps. In such systems, at least on-site Hubbard repulsion can be tuned to realize AFM order, as recently demonstrated on various Euclidean lattices~\cite{EsslingerReview2010, EsslingerPhysRevLett2013, ZwierleinScience2016, MarkusGreinerNature2017}.

\emph{Acknowledgments}.~N.G.\ was supported by a Startup grant of B.R. from Lehigh University. S.M.\ was supported by Weizmann Institute of Science, Israel Deans fellowship through Feinberg Graduate School and Raymond and Beverly Sackler Center for Computational Molecular and Material Science at Tel Aviv University. B.R.\ was supported by NSF CAREER Grant No.\ DMR-2238679.

\bibliography{HyperbolicReferences}

\begin{thebibliography}{45}
\bibitem [{com()}]{comment1}%
  \BibitemOpen
  \href@noop {} {\bibinfo  {journal} {Sites belonging to the center $p$-gon
  constitute the zeroth generation and each successive layers of sites
  constitute the progressively next generations of hyperbolic lattices. The
  sites from the zeroth generation or center $p$-gon of the system is also
  coined `center' in Figs.~1 and 2}\ }\BibitemShut {NoStop}%
\end{thebibliography}

\end{document}